\documentclass{caosp310}

\usepackage{graphicx}
\usepackage{amsmath}
\usepackage{natbib}
\articleNo{ }
\pubyear{ }
\volume{ }
\volnumber{ }
\firstpage{ }
\received{September 26, 2025}
\accepted{November 20, 2025}

\def\BibTeX{{\rm B\kern-.05em{\sc i\kern-.025em b}\kern-.08em
             T\kern-.1667em\lower.7ex\hbox{E}\kern-.125emX}}

\begin{document}

%
\hauthor{J.U.\,Guerrero-González, V.\,Chavushyan and V.M.\,Patiño-Álvarez}

\title{Exploring Emission Line Variability and Jet-Broad Line Region Interaction in the Blazar TON 599}


%
%
\author{
  Jonhatan\,U.\,Guerrero-Gonz\'alez\inst{1}\orcid{0009-0002-2044-3274} 
  \and
  Vahram\,Chavushyan\inst{1}\orcid{0000-0002-2558-0967}
  \and
  Víctor\,M.\,Pati\~no-Álvarez\inst{1,2}\orcid{0000-0002-5442-818X}
}

%
\institute{
  Instituto Nacional de Astrof\'isica, \'Optica y Electr\'onica (INAOE), 
  Luis Enrique Erro \#1, Tonantzintla, Puebla, M\'exico, C.P.\,72840
  \and
  Max-Planck-Institut f\"ur Radioastronomie, Auf dem H\"ugel 69, D-53121 Bonn, Germany
}

\date{March 8, 2003}
\maketitle

\begin{abstract}
Blazars, a highly variable Active Galactic Nuclei (AGNs) subclass, provide a unique opportunity to explore the physical processes within their relativistic jets and emission regions. In this study, we investigate the multiwavelength variability of the blazar TON 599, a Flat Spectrum Radio Quasar (FSRQ), with a particular emphasis on its emission line behavior. We focus on the Mg II $\lambda$2798 Å emission line, a key tracer of the ionized gas in the broad-line region (BLR), and its role in jet-induced variability. In addition to optical emission lines, we analyze gamma-rays (0.1–300 GeV), X-rays (0.2–10 keV), optical continuum ($\lambda$3000 Å), optical polarization, and millimeter-wavelength light curves. Three cross-correlation methods are employed to investigate temporal relationships between the emission line and continuum across various wavelengths. Using the Non-Thermal Dominance (NTD) parameter, our analysis confirms that synchrotron emission dominates the continuum during active states, highlighting the jet's primary role in the observed variability. The Mg II emission line exhibits quasi-simultaneous variability with the optical continuum, suggesting photoionization driven by the jet's non-thermal radiation. Additionally, the minimal time lag between gamma-ray and optical/near-ultraviolet emissions supports a synchrotron self-Compton origin for the most variable component of the gamma-ray emission. These findings highlight the importance of emission line variability and multiwavelength observations in constraining the interaction between jets and the BLR in blazars. The results contribute to a deeper understanding of AGN emission mechanisms and the complex interplay between jets and their surrounding environments.
\end{abstract}

%

\section{Introduction}

Supermassive black holes (SMBHs) reside at the centers of most galaxies, and those with high accretion rates give rise to Active Galactic Nuclei (AGN) \citep{Urry1995}. A remarkable subclass of AGN is represented by blazars, whose relativistic jets are aligned close to our line of sight. They are powerful radio sources, characterized by extreme variability across the entire electromagnetic spectrum, with timescales ranging from years to minutes. Blazars provide a unique laboratory for studying physical conditions that cannot be reproduced on Earth.

Blazars are commonly divided into two subclasses: BL Lacertae objects (BL Lacs) and Flat Spectrum Radio Quasars (FSRQs). In particular, FSRQs tend to show broad optical emission lines (e.g., H$\beta$, Mg II, C IV) in addition to a non-thermal continuum \citep{Urry1995,Veron2000}. By contrast, BL Lacs are characterized by optical spectra dominated by a nearly featureless non-thermal continuum with very weak or absent emission lines \citep{Stickel1991}.

Within this context, this work focuses on the blazar TON 599 ($z = 0.725$; \citealt{Hewett2010}), classified as an FSRQ. We investigate its multifrequency variability, including gamma rays (0.1--300 GeV), X-rays (0.2--10 keV), the UV spectral continuum (3000 \AA), the Mg II $\lambda2798$ \AA{} emission line, optical polarization (5000--7000 \AA), and 1 mm emissions. Cross-correlation analysis is employed to explore connections among these bands. This study provides new insights into the multifrequency emission processes of TON 599 and contributes to a broader understanding of blazar-type AGN.

\section{Object of Study}

The blazar TON 599 is classified as a Flat Spectrum Radio Quasar (FSRQ) due to its strong broad line emission and small viewing angle, making it one of the most variable types of AGN \citep{Wills1983, Prince2019, Hallum2022}.  

Originally discovered in the 1950s by \citet{Iriarte}, TON 599 was included in a catalog of more than 800 ``blue stars'' identified near the North Galactic Pole, as part of an observational program carried out with the Tonantzintla Schmidt Camera, in Puebla, Mexico. Since then, this source has been extensively monitored by multiple facilities, yielding a rich multi-band archive and numerous publications that consistently confirm its strong variability across different wavelengths.  

\begin{figure}[ht!]
\centering
\begin{center}
\includegraphics[width=1\textwidth]{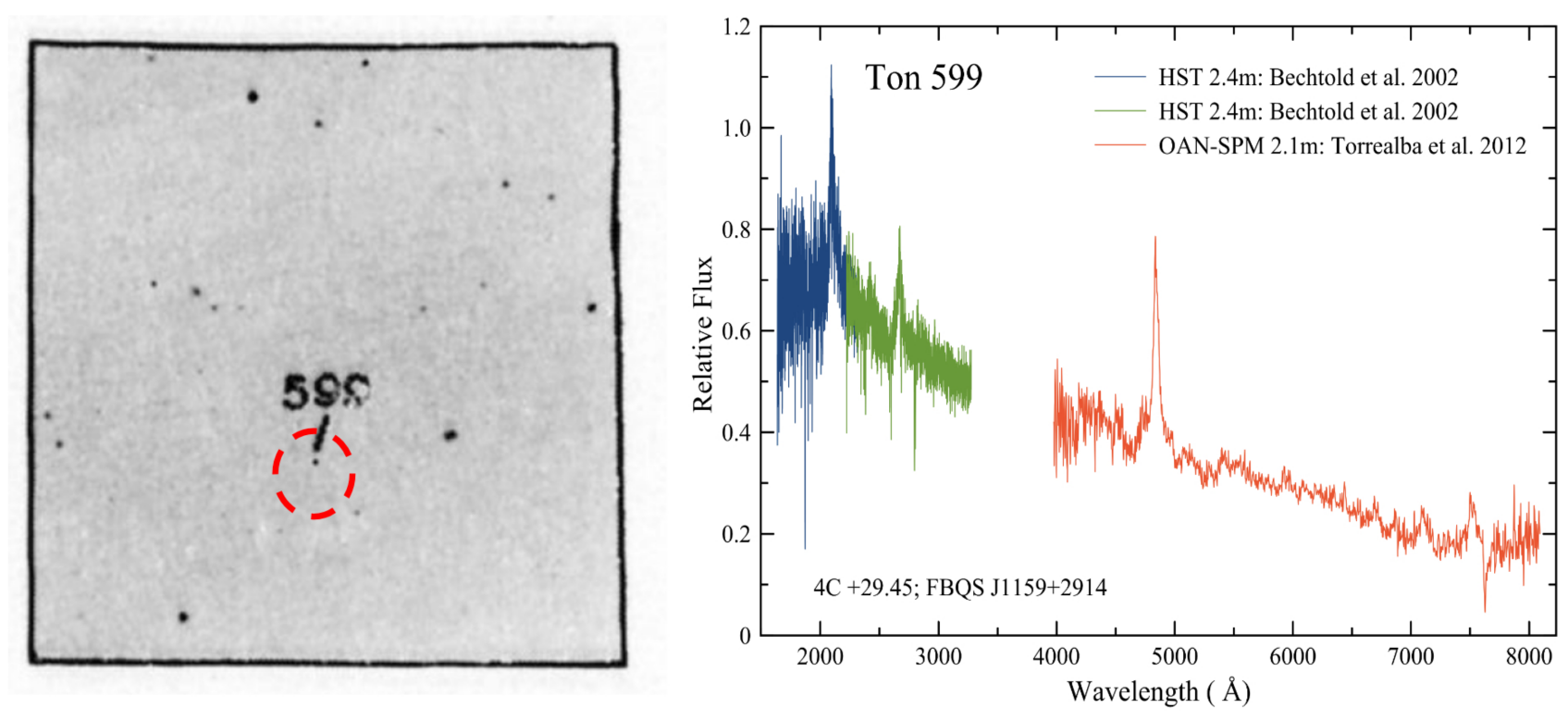} 
\caption{Left: Identification chart of TON 599 \citep{Iriarte}. Right: Observed frame UV/Optical spectrum of TON 599 \citep{Bechtold2002, Torrealba2012}}
\label{Fig1}
\end{center}
\end{figure}

Figure~\ref{Fig1} shows the original identification chart of TON 599 together with its UV/optical spectrum, illustrating both the historical context of its discovery and the spectral features that define it as an FSRQ.

\section{Methodology}

The observations analyzed in this work span multiple wavelengths. Gamma-ray data were obtained from \textit{Fermi}-LAT, while X-ray data were taken from the \textit{Swift-XRT Monitoring of Fermi-LAT Sources of Interest}. Optical spectra were retrieved from the database of the Ground-Based Observational Program associated with the \textit{Fermi} Gamma-ray Space Telescope at the University of Arizona, and 1 mm observations were collected from the Submillimeter Array (SMA) database.  

\begin{figure}[ht!]
\centering
\begin{center}
\includegraphics[width=0.63\textwidth]{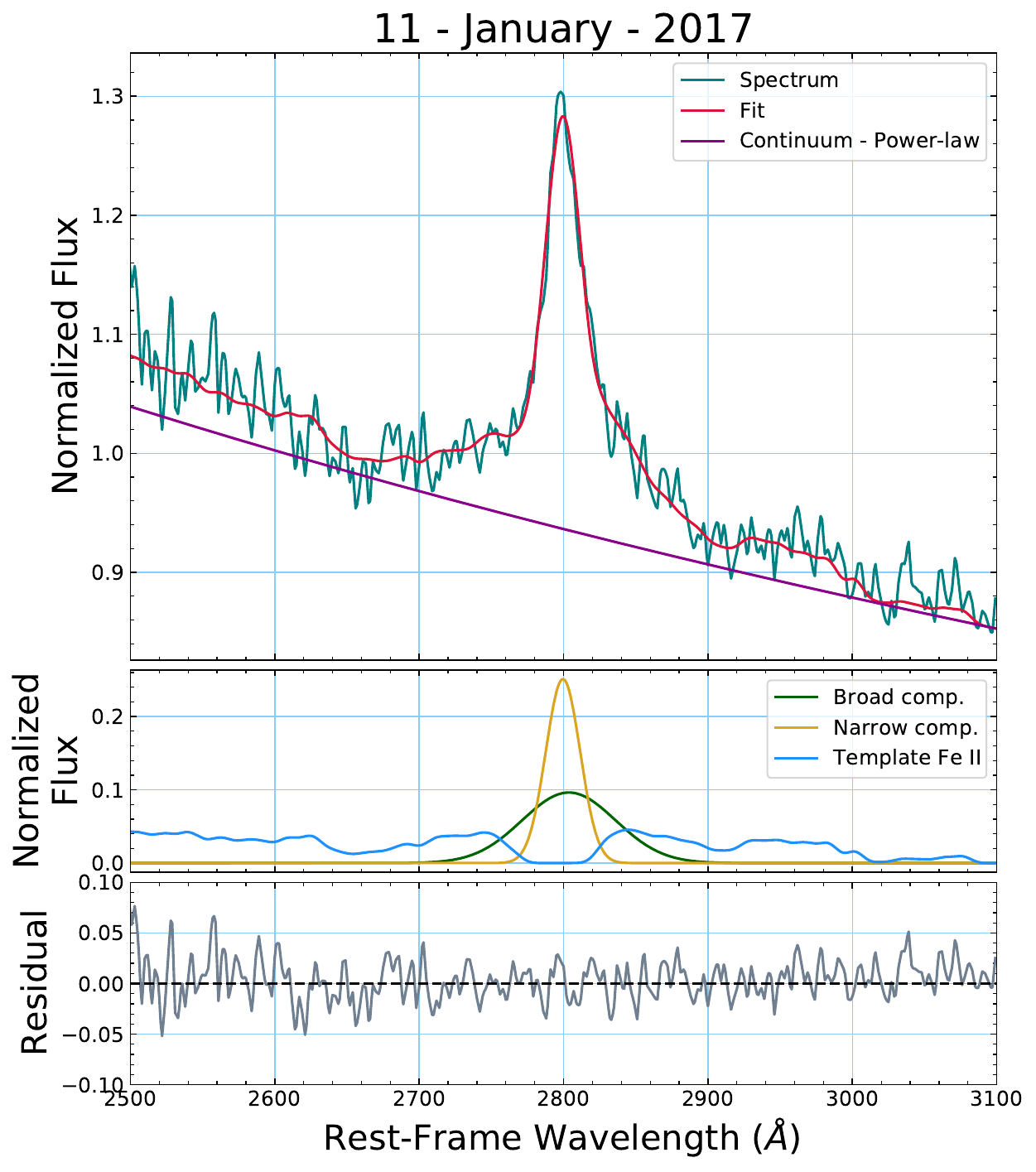} 
\caption{Example of a TON 599 spectrum (January 2017, Steward Observatory) showing the multi-component fit and residuals.}
\label{Fig2}
\end{center}
\end{figure}

In total, 196 spectra of TON 599 were compiled, of which two were discarded due to low signal-to-noise ratio. For each remaining spectrum, a fitting procedure was performed on the Mg II $\lambda2798$ \AA{} emission line using multi-component models. The continuum was fitted with a power-law function, the Fe II emission in the UV was modeled with empirical templates \citep{Vestergaard2001}, and the Mg II line itself was decomposed into broad and narrow components with Gaussian profiles (see Figure~\ref{Fig2} for an example of the fitting procedure). This approach allowed us to reliably characterize the spectral features of TON 599. 

From these fits we derived integrated fluxes, line and continuum luminosities, and subsequently physical parameters such as the Non-Thermal Dominance (NTD), which are discussed in the following section.

\section{Non-Thermal Dominance Parameter: NTD}

The non-thermal dominance (NTD) parameter, introduced by \citet{Shaw2012}, quantifies the contribution of the relativistic jet to the optical/UV continuum and is defined as:

\begin{equation}
    NTD = \frac{L_{obs}}{L_{pred}},
\end{equation}

where $L_{obs}$ is the observed luminosity and $L_{pred}$ is the predicted luminosity inferred from emission lines in non-blazar AGNs. Since broad emission lines trace the accretion disk power (i.e. $L_{pred}$=$L_{disk}$), \citet{Patino2016} reformulated NTD as:  

\begin{equation}
    NTD = 1 + \frac{L_{jet}}{L_{disk}}, \;\;\; NTD \geq 1,
\end{equation}

which allows different regimes to be identified: $NTD=1$ (only disk emission), $1<NTD<2$ (disk-dominated), $NTD=2$ (equal contributions from disk and jet), and $NTD>2$ (jet-dominated).  

The value of $L_{pred}$ was obtained from the empirical relation between $L_{MgII}$ and $L_{3000\,\text{Å}}$ derived from a sample of $\sim$ 40,000 radio-quiet AGN \citep{patino2025} :

\begin{equation}
    \log L_{MgII} = (0.826 \pm 0.025) \log \lambda L_{\lambda3000} + (6.057 \pm 1.164).
    \label{eq2.19}
\end{equation}

Here, $L_{MgII}$ was measured from the integrated flux of the emission line, while $L_{obs}$ was derived from the continuum flux at 3000~Å, both calculated using a luminosity distance of $D_{L} = 4445.5$ Mpc ($z=0.725$).

In addition, single-epoch black hole mass estimates were obtained using standard virial relations based on the Mg II emission line luminosity and the 3000\,Å continuum \citep{Kong2006, Vestergaard2006, Shen2011}. These methods assume that the Broad Line Region (BLR) is virialized and that the continuum is primarily powered by the accretion disk.  

\begin{figure}[ht!]
\centering
\begin{center}
\includegraphics[width=0.92\textwidth]{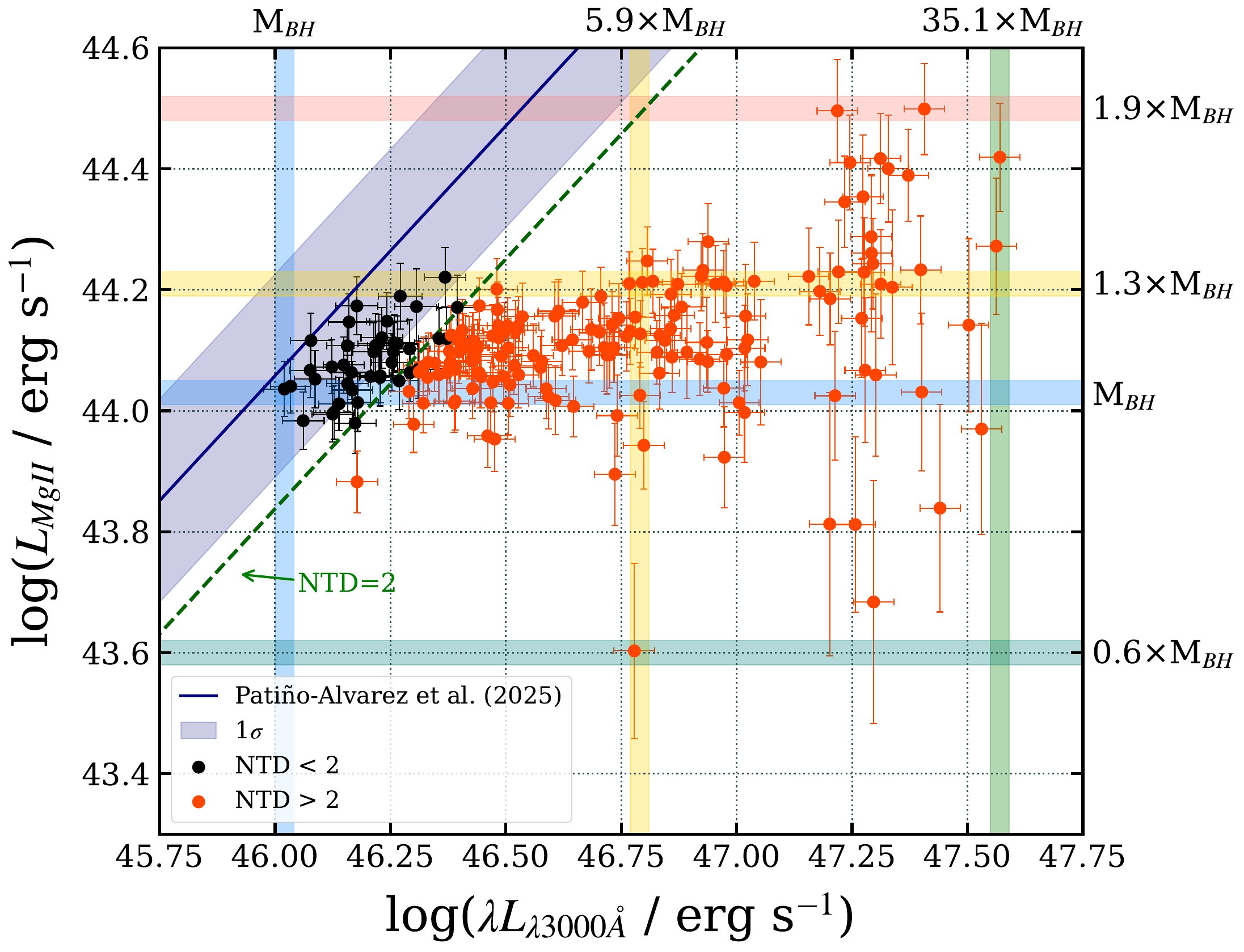} 
\caption{Variations in the Mg II emission line luminosity versus the continuum luminosity at 3000 Å for TON 599. The right and top axes show overestimations and underestimations of the black hole mass when calculated using single-epoch techniques, using the continuum and emission line luminosity, respectively.}
\label{Fig3}
\end{center}
\end{figure}

For jet-dominated sources such as TON 599, however, the continuum can be significantly contaminated by non-thermal emission, which may lead to systematic overestimations of $M_{\mathrm{BH}}$. Since the Mg II emission line is expected to be less affected by this contamination, mass estimates derived from $L_{MgII}$ are considered more reliable in these cases. Figure~\ref{Fig3} shows the relation between $L_{MgII}$ and $L_{3000\,\text{Å}}$ for TON 599, where deviations from the empirical trend directly reflect the increasing jet contribution and its impact on black hole mass determinations.

\section{Cross-Correlation Analysis}

To investigate possible time lags between the continuum and emission line variability, we applied cross-correlation functions (CCFs) using three complementary approaches: the interpolated CCF \citep{Gaskell1986}, the discrete CCF \citep{Edelson1988}, and the Z-transformed DCF \citep{Alexander1997}, all incorporating non-stationarity corrections \citep{Patino2013, Amaya-Almazan2022}.  

The statistical significance of the correlations was evaluated through Monte Carlo simulations \citep{Timmer1995, Emmanoulopoulos2013}, with confidence levels at 90, 95, and 99\% (grey lines in Figure 4). In this work, we adopt the 99\% level as the threshold for reliable correlations. Employing multiple CCF methods enhances robustness by mitigating the effects of interpolation gaps, spurious peaks, and aliasing, among other caveats the data might present.

\begin{figure}[ht!]
\centering
\begin{center}
\includegraphics[width=1\textwidth]{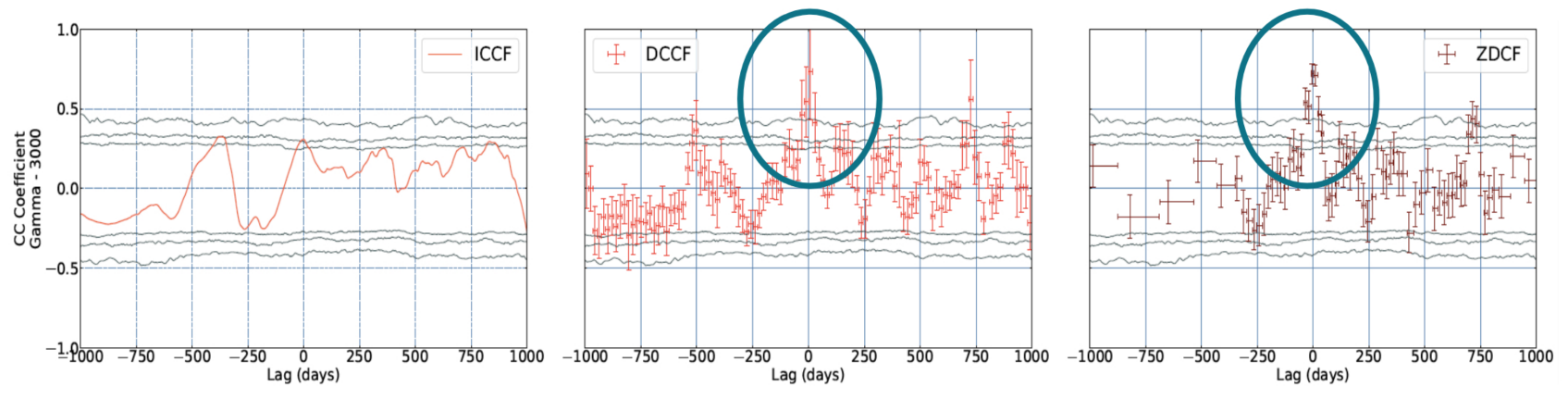} 
\caption{Example of the cross-correlation functions; in this case between the gamma-ray and 3000 Å light curves using three methods. Left: interpolated cross-correlation function (ICCF); middle: discrete cross-correlation function (DCCF); right: Z-transformed discrete correlation function (ZDCF).}
\label{Fig4}
\end{center}
\end{figure}

As an illustrative example, Figure~\ref{Fig4} presents the results of the cross-correlation between the gamma-ray and 3000~Å light curves using the three methods, illustrating both the consistency of the techniques and the regions where significant correlations are detected. In this figure, we see that the ICCF does not exceed the 99\% confidence level, whereas the DCCF and ZDCF show a significant correlation within the same lag range. This difference stems from the ICCF’s sensitivity to interpolation and irregular sampling. Therefore, for this specific case we consider the detection supported by DCCF/ZDCF to be more reliable.

\clearpage

\section{Results}

The results presented below are derived from a joint analysis of the multi-frequency light curves and the cross-correlations between bands. In particular, the cross-correlations provide information on the time lags between different wavelength bands of the electromagnetic spectrum, which may be related to distances between emitting regions, provided that factors such as the jet viewing angle and the wavelengths involved are taken into account.

\begin{figure}[ht!]
\centering
\begin{center}
\includegraphics[width=1.15\textwidth]{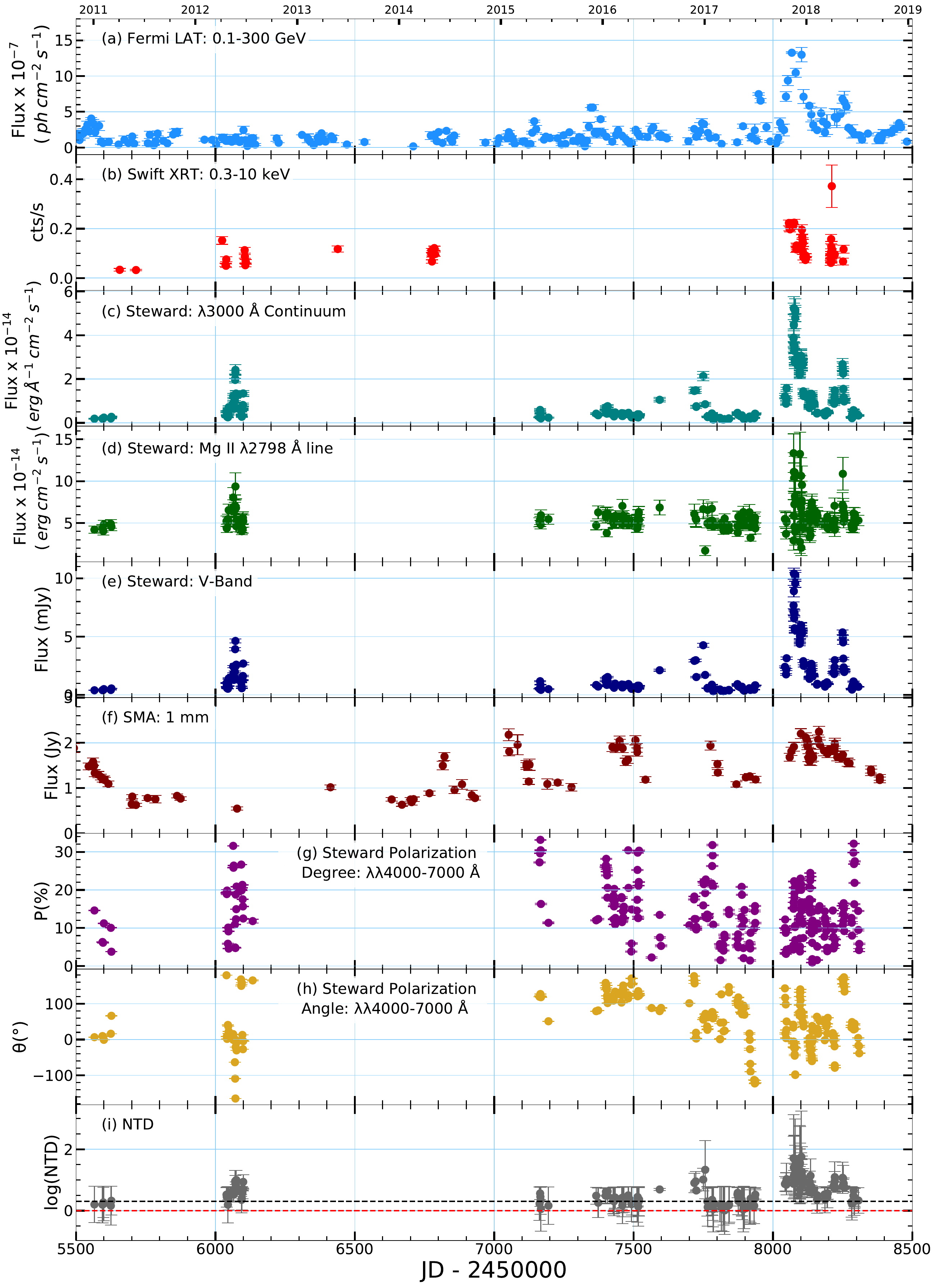} 
\caption{Multifrequency Light Curves for TON 599.}
\label{Fig5}
\end{center}
\end{figure}

Figure~\ref{Fig5} shows the multi-frequency light curves of TON~599, including \textit{Fermi}-LAT $\gamma$ rays (0.1--300\,GeV), \textit{Swift}-XRT X rays (0.3--10\,keV), optical data from the \textit{Steward Observatory} (4000--7000\,\AA\ in the observer’s frame), comprising the rest-frame 3000\,\AA\ continuum, the Mg II $\lambda2798$\,\AA\ line flux, $V$-band photometry, the polarization degree, and the polarization angle (measured between 5000 and 7000\,\AA), SMA 1\,mm emission, and the Non-Thermal Dominance (NTD) curve computed from the 3000\,\AA\ continuum and the Mg II line.

Consistent with the cross-correlation analysis, the $V$ band and the 3000\,\AA\ continuum vary quasi-simultaneously, with contemporaneous rises and declines in Figure~\ref{Fig5} (panels c,e). The NTD in panel (i) increases over the same intervals to values above 2 (black dashed line), indicating enhanced jet contribution. The $\gamma$-ray activity (panel a) rises during the main optical/NUV events, in agreement with the significant correlations recovered by the discrete and $Z$-transformed discrete correlation functions. The Mg II $\lambda2798$\,\AA\ emission line (panel d) exhibits flux enhancements that coincide with optical/NUV high states, consistent with near-zero lags found for the line–continuum pairs.

By contrast, some bands do not show clear correlations. The X-ray coverage (panel b) is limited and does not sample the strongest optical/NUV outbursts, which restricts any association. The 1\,mm emission (panel f) traces longer variability timescales and lacks dense sampling during the short optical/NUV flares, reducing the expected cross-correlation signal. The polarization degree and polarization angle (panels g--h) show variability but no consistent temporal relation with the other bands in this dataset. Finally, the $\gamma$-ray–Mg II correlation is inconclusive, mainly due to cadence and sampling differences between the high-energy and emission line light curves.

\clearpage

\section{Conclusions}

The cross-correlation analysis and multifrequency monitoring of TON 599 allow us to identify several key results regarding the connection between the jet, the optical/UV continuum, the emission lines, and the high-energy emission. These findings can be summarized as follows:

\begin{enumerate}
    \item The cross-correlations between the $V$ band, the 3000 \AA\ continuum, and the NTD are consistent with zero lag. This indicates co-spatiality among the optical/NUV emissions, with the main variability in NTD arising from the jet. The similar variability patterns observed in both the $V$ band and the 3000 \AA\ continuum support the interpretation that these bands are dominated by synchrotron radiation from the jet.

    \item The correlations between the optical/NUV bands and gamma rays suggest that the seed photon population is dominated by synchrotron emission, and that the gamma-ray variability is mainly produced by the Synchrotron Self-Compton (SSC) process during high states of the source. This finding that SSC dominates challenges the standard external Compton scenario and has implications for jet models and Spectral Energy Distribution (SED) studies.

    \item Cross-correlations between the optical/NUV bands and the Mg II $\lambda2798$ \AA\ emission line also yield a zero lag. The temporal coincidence of line flares with optical/NUV variability implies that the line-emitting gas is being ionized by synchrotron radiation from the jet, in addition to the accretion disk.

    \item The observed relation between continuum and line luminosities deviates from expectations for a radio-quiet BLR, indicating that the emission line is not solely ionized by the accretion disk. As a result, black hole mass estimates based on single-epoch methods should preferentially use epochs when the continuum is disk-dominated to obtain more realistic values.
\end{enumerate}


Overall, TON~599 demonstrates that optical/NUV synchrotron emission drives both $\gamma$–ray SSC variability and Mg II ionization, revealing a strong interplay between the relativistic jet and the BLR. This challenges canonical reverberation mapping assumptions and adds further evidence, consistent with sources such as 3C~454.3 \citep{LeonTavares2013, Amaya-Almazan2021}, 3C~279 \citep{Patino2018}, and CTA~102 \citep{Chavushyan2020}, that BLR gas can be directly influenced by jets.  

The case of TON~599 highlights how jet–BLR interactions significantly impact both variability patterns and black hole mass determinations in FSRQs. Future efforts should focus on short-timescale flares, testing the $NTD$ threshold across different bands, refining Mg II-based mass estimates with improved instrumental corrections, and extending this methodology to a larger sample of blazars \citep{Massaro}.

\clearpage

\acknowledgements
This work has been supported by the SECIHTI program during my doctoral studies. Support was also provided by the MPIfR-Mexico Max Planck Partner Group, led by Dr. V\'ictor Manuel Pati\~no \'Alvarez, whose significant contribution to this research is greatly appreciated.

\bibliography{References}

\clearpage

\end{document}